\documentclass[prb,aps,amssymb,shownopacs,twocolumn,superscriptaddress,times,10pt,notitlepage]{revtex4-1}
\usepackage{amsmath}
\usepackage{amssymb}
\usepackage{amsfonts}
\usepackage{listings}
\usepackage{enumerate}
\usepackage{latexsym}
\usepackage{bm}
\usepackage{graphicx}
\usepackage{braket}
\usepackage[pdftex,colorlinks=false]{hyperref}
\usepackage{xcolor}
\usepackage{verbatim}
\usepackage{float}
\usepackage{multirow}
\usepackage{MnSymbol}
\usepackage[caption=false]{subfig}

\usepackage{times}

\setcitestyle{numbers,square}

\begin{document}

\tolerance 10000

\newcommand{\cbl}[1]{\color{blue} #1 \color{black}}

\newcommand{\vk}{{\bf k}}

\widowpenalty10000
\clubpenalty10000

\title{Study of the Two-Dimensional Frustrated J1-J2 Model with Neural Network Quantum States}

\author{Kenny Choo}
\address{
Department of Physics, University of Zurich, Winterthurerstrasse 190, 8057 Zurich, Switzerland
}

\author{Titus Neupert}
\address{
 Department of Physics, University of Zurich, Winterthurerstrasse 190, 8057 Zurich, Switzerland
}

\author{Giuseppe Carleo}
\address{
Center for Computational Quantum Physics, Flatiron Institute, 162 5th Avenue, New York, NY 10010, USA
}

\begin{abstract}
The use of artificial neural networks to represent quantum wave-functions has recently attracted interest as a way to solve complex many-body problems. The potential of these variational parameterizations has been supported by analytical and numerical evidence in controlled benchmarks. While approaching the end of the early research phase in this field, it becomes increasingly important to show how neural-network states perform for models and physical problems that constitute a clear open challenge for other many-body computational methods. In this paper we start addressing this aspect, concentrating on a presently unsolved model describing two-dimensional frustrated magnets. Using a fully convolutional neural network model as a variational ansatz, we study the frustrated spin-1/2 $J_1$-$J_2$ Heisenberg model on the square lattice. We demonstrate that the resulting predictions for both ground-state energies and properties are competitive with, and often improve upon, existing state-of-the-art methods. 
In a relatively small region in the parameter space, corresponding to the maximally frustrated regime, our ansatz exhibits comparatively good but not best performance. The gap between the complexity of the models adopted here and those routinely adopted in deep learning applications is, however, still substantial, such that further improvements in future generations of neural-network quantum states are likely to be expected. 
\end{abstract}

\date{\today}

\maketitle

\section{Introduction}
With the ever-improving computational resources, techniques, and datasets machine learning has in recent years proven itself to be an extremely versatile tool for solving tasks previously thought impossible for a computer in the near future \cite{Silver2017}. Deep learning, a branch of machine learning based on the use of deep artificial neural networks (ANN) has played a very important role in these developments, and it is currently largely believed that it will be an important computational method for years to come. In the field of condensed matter physics, several machine learning applications have been put forward in recent years. For example, they have been used to successfully classify phases of matter~\cite{Carrasquilla2017,Nieuwenburg2017,Schindler2017, Venderley2018}, to perform quantum state tomography~\cite{Torlai2018, rocchetto_learning_2018, Carrasquilla2018}, to simulate quantum computers~\cite{Jonsson2018}, to classify experimental data~\cite{Zhang2018,rem_identifying_2018,bohrdt_classifying_2018}, and much more.

In the realm of computational quantum physics, the restricted Boltzmann machine (RBM), a type of ANN, was proposed as a variational ansatz~\cite{Carleo2017} for many-body quantum systems. Since then, there has been a burst of research investigating the viability of such ansatz. It has been shown that unlike other variational ansatz~\cite{White1992}, wavefunctions based on RBMs can potentially capture long-ranged and volume law scaling entanglement~\cite{Deng2017}. Its representation properties have also been extensively characterized, and it by now known that RBM and related states can efficiently describe the ground states of many physical Hamiltonians~\cite{Gao2017, Nomura2017, Chen2018, Glasser2018}. Motivated by these theoretical successes, the community has pushed ahead and explored a wide variety of ANNs such as feedforward neural networks~\cite{Saito2017, Saito2018, Choo2018, Liang2018}, deep Boltzmann machines~\cite{Carleo2018, Pastori2018}, and a variety of other neural-network inspired ansatz~\cite{Luo2018, Kochkov2018}. These approaches have undergone an extensive phase of benchmarks, and have been compared to existing exact results, both in one and two dimensions, typically showing very good accuracy for ground and excited-state properties. While this phase of benchmarks has been overall important to assess the potential of this approach, the method is yet to be fully deployed on manifestly open problems, for which the application of ANN states could prove beneficial to resolve inconclusive results from other methods.  

Here, we consider the case of the antiferromagnetic spin-1/2 $J_1$-$J_2$ model on the square lattice, a prototypical frustrated magnetic system for which no exact solution is known. Despite active research on the model for the past few decades, one of the chiefly open questions is whether a spin liquid phase exists around the point of maximum frustration. Numerous computational methods \cite{Zhitomirsky1996,Capriotti2000,Mambrini2006,JFYu2012, Sachdev1990,Chubukov1991,Singh1999, Capriotti2001,GMZhang2003,HCJiang2012,LWang2013,WJHu2013,SSGong2014,Richter2015, Morita2015, Liu2018} have been used to study this problem, often finding conflicting conclusions. The active and long-lasting interest in the $J_1$-$J_2$ model makes it an especially useful and non-trivial benchmark for ANNs techniques, since the best of the currently available many-body techniques have been used to study it. 

In this paper we demonstrate that ANNs are a viable and competitive variational ansatz to study frustrated spin models in two dimensions.  Our wave-function is parametrized as a type of feedforward neural network known as a convolutional neural network (CNN), routinely used in top applications of deep learning in industry. We concentrate on the two-dimensional $J_1$-$J_2$ model on the square lattice, providing results for both the ground-state energy and spin correlation functions. We benchmark our results on the $6\times6$ cluster with exact diagonalization results \cite{Schulz1996} and on the $10\times10$ cluster with density matrix renormalization group (DMRG) calculations~\cite{SSGong2014} as well as traditional variational Monte Carlo (VMC) based on Gutzwiller-projected mean field Fermionic wavefunctions \cite{WJHu2013}. We show that for several points in the phase diagram the CNN variational energies we obtain improve upon those obtained by the other techniques. We also find that there is a small window of frustration ratios for which our method, while being competitive with the state of the art, is not yet delivering cutting-edge results. Finally, we discuss the origin of these limitations and some strategy to further improve ANN-based methods in future works. 

\begin{figure*}[t]
            \includegraphics[width=1.0\textwidth]{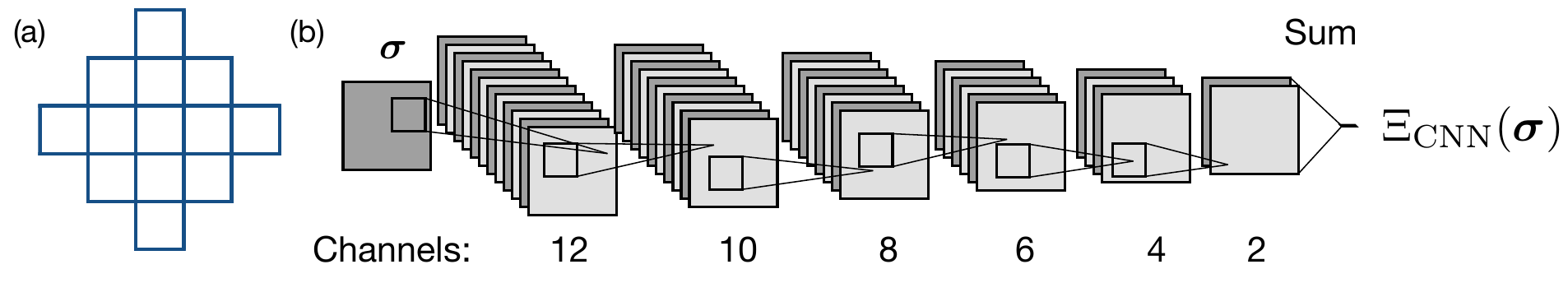}
    \caption{Network architecture. (a) Shape of the filter which we apply across the square lattice. (b) Full architecture of the convolutional neural network used in this work. There are $6$ convolutional layers followed by an output layer which simply sums the values of the prenultimate layer. In the first layer, we use the logarithm of the hyperbolic cosine as the activation function $g_{\textrm{lncosh}}(z) = \log [\cosh(z)]$, while in all other layers the activation function is given by the complex generalization of the ReLU \cite{Trabelsi2018}. The total number of complex-valued parameters in the network is $3838$.}
\label{fig: CNN}
\end{figure*}

\section{Model}
The spin-1/2 $J_1$-$J_2$ Heisenberg model is defined by the Hamiltonian
\begin{equation}\label{eq: J1J2}
\hat{H} = J_{1}\sum_{\langle ij \rangle} \hat{\boldsymbol{S}}_{i} \cdot \hat{\boldsymbol{S}}_{j} + J_{2}\sum_{\llangle ij \rrangle}  \hat{\boldsymbol{S}}_{i} \cdot \hat{\boldsymbol{S}}_{j},
\end{equation}
where $\hat{\boldsymbol{S}}_{i} = (\hat{S}_{i}^{x}, \hat{S}_{i}^{y}, \hat{S}_{i}^{z})$ representing here the spin operators at site $i$ of a square lattice with periodic boundary conditions. The symbols $\langle \cdots \rangle$ and $\llangle \cdots \rrangle$ indicate pairs of nearest and next-nearest neighbor sites, respectively. We are interested in the case where both the nearest and next-nearest neighbor interactions are anti-ferromagnetic, i.e., $J_{1}, J_{2} \geq 0$, so that the magnetic interactions are frustrated. For simplicity, we fix $J_{1} = 1$ throughout this paper. In addition, since total magnetization is a conserved quantity in this model and it is expected that the ground state is in the zero magnetization sector, we shall restrict our analysis to this sector, denoting by $\boldsymbol{\sigma}$ spin configurations belonging to it.

When $J_{2} = 0$, the system is unfrustrated and there is a well established Neel long range order \cite{Sandvik1997,Buonaura1998}. In the opposite limit $J_{2}\gg J_1$, the system is also magnetically ordered with pitch vector $\boldsymbol{Q} = (\pi, 0)$ or $(0,\pi)$. However, in the intermediate regime, where $J_{2}/J_{1} \approx 0.5$, the system is highly frustrated. There have been numerous conflicting proposals for the candidate ground state, such as the plaquette valence-bond state \cite{Zhitomirsky1996,Capriotti2000,Mambrini2006,JFYu2012}, the columnar valence-bond state \cite{Sachdev1990,Chubukov1991,Singh1999} or a gapless spin liquid \cite{Capriotti2001,GMZhang2003,LWang2013,WJHu2013,Richter2015, SSGong2014, Liu2018}, but the correct answer is still unknown.

While we make no claim about the nature of the physics in the frustrated regime $J_{2}/J_{1} \approx 0.5$, we hope to present a new and potentially viable ansatz for variational methods. We shall compare our variational energies with other methods such as DMRG and traditional VMC based on a projected fermionic ansatz and show that there are regimes in parameter space, where neural network states have the lowest variational energies. 

\section{Neural Network Quantum States}
Neural network quantum states (NQS) were first proposed in Ref.~\onlinecite{Carleo2017}, where a RBM was used as a variational ansatz for the Heisenberg model on a square lattice, corresponding to $J_{2}=0$ in Eq.~\eqref{eq: J1J2}.  Since neural networks are essentially functions with a large number of parameters, the idea behind NQS is to interpret the output of a network as the complex amplitudes of a wavefunction $\Psi(\boldsymbol{\sigma})$, where $\boldsymbol{\sigma}$ is a vector representing a spin configuration of the system. In this paper, we shall use a variational wavefunction that is expressed as a feedforward CNN, which has been shown to be able to support volume law entanglement more efficiently than the RBM~\cite{Yoav2019}.

A feedforward neural network has the most basic structure of a variational function composed of a series of transformations called ``layers". It takes an input vector $\boldsymbol{v}^{(0)}$ and successively applies a sequence of layers to map it to the output, i.e., $\boldsymbol{v}^{(0)} \rightarrow \boldsymbol{v}^{(1)} \rightarrow \cdots \rightarrow \boldsymbol{v}^{(L)}$ for a $L$ layer network. A generic layer implements an affine transformation followed by a non-linear transformation which is usually taken to be a coefficient wise operation
\begin{equation}\label{eq: fullyconnected}
\boldsymbol{v}^{(n)}_{i} = g\left(\sum_{j} W_{ij} v^{(n-1)}_{j} + b_{i}\right),
\end{equation}
where $W_{ij}$ are the elements of a weight matrix, $ \boldsymbol{b}_{i}$ are the elements of a bias vector, and $g$ is some non-linear function. Since wavefunctions are in general complex-valued, we use complex-valued weights and biases. The non-linear function is then a function over the complex numbers, i.e., $g: \mathbb{C} \to \mathbb{C}$. For the simulations done in this work, $g$ is either the rectified linear unit (ReLU) generalized to complex numbers~\cite{Trabelsi2018}
\begin{equation}
g_{\mathrm{relu}}(z) =
\begin{cases}
z & \textrm{if } \frac{3\pi}{4} > \arg{z} \geq -\frac{\pi}{4} \\
0 & \textrm{otherwise}
\end{cases}
\end{equation}
or the logarithm of the hyperbolic cosine $g_{\textrm{lncosh}}(z) = \log [\cosh(z)]$.

In a CNN, the layers have a spatially local structure and are called convolutional layers. 
A convolutional layer is also an affine transformation as in Eq.~\eqref{eq: fullyconnected}, but with certain constraints being placed on the weights and biases: The transformation is separated into several independent ``channels'' and each channel is characterized by a ``filter'' matrix. A convolutional layer  indexed by $n$ with $C_{n-1}$ input channels and $C_{n}$ output channels will perform the transformation 
\begin{equation}
\boldsymbol{v}^{(n)}_{m,j} = g \left( \sum_{l=1}^{C_{n-1}} \sum_{k=1}^{\mathcal{K}^{(n)}} K^{(n)}_{m,k} \boldsymbol{v}^{(n-1)}_{l,a_{jk}} + b^{(n)}_{m} \right),
\end{equation}
where $m,l$ are channel indices, $K^{(n)}_{m,k}$ are filter parameters in channel $m$ and $b^{(n)}_{m}$ is bias in channel $m$. The index $a_{jk}$ indicates the position of the input image to be acted on by the $k$ parameter of the filter so as to contribute to the $j$ position of the output vector.
The complete structure of our CNN and shape of the convolutional filters are shown in Fig.~\ref{fig: CNN}. Notice that in the final layer, all outputs of all channels are summed over. This constitutes a so-called average pooling layer. The full network has a total of $3838$ complex-valued parameters, independent of the system size that we will apply it to.

The complete CNN thus represents an explicitly translationally invariant function with zero momentum defined on the configurations of the Hilbert space, i.e., $\Xi_{\mathrm{CNN}}: \lbrace \boldsymbol{\sigma} \rbrace \to \mathbb{C}$ where $\boldsymbol{\sigma}$ refers to a computational basis of the Hilbert space. The variational wavefunction represented by the CNN is then given by
\begin{equation} \label{eq: bare CNN}
\Psi_{\mathrm{CNN}} (\boldsymbol{\sigma}) = \exp\left[\Xi_{\mathrm{CNN}}(\boldsymbol{\sigma})\right].
\end{equation}
One of the advantages using such a fully convolutional structure resides in an intrinsically more efficient learning procedure, as opposed to fully connected layers. For example, the set of kernels optimized for a smaller system size provides a good starting point for the optimization of a larger system size such that relatively few iterations is needed for convergence. We use the same network structure, with the same number of variational parameters, for all system sizes studied.

\subsection*{Sign structure of the ground state}

It is known that in the extremal limits ($J_{1}=0$ or $J_{2}=0$) the ground state wavefunction obeys a simple sign rule. In those limits, the ground state wavefunction takes the form
\begin{equation}
\ket{\mathrm{GS}} = \sum_{\boldsymbol{\sigma}} (-1)^{M_{\mathcal{A}}(\boldsymbol{\sigma})} \Psi(\boldsymbol{\sigma}) \ket{\boldsymbol{\sigma}}
\end{equation}
where $M_{\mathcal{A}}(\boldsymbol{\sigma})$ is the total number of up spins on a subset $\mathcal{A}$ of the sites and $\Psi(\boldsymbol{\sigma}) \geq 0$. When $J_{2}=0$, the subset $\mathcal{A}$ is given by one of the two bipartite components of the square lattice, leading to the so-called Marshall-Peierls sign rule~\cite{Marshall1955}. On the other hand, when $J_{1} = 0$, $\mathcal{A}$ can be chosen either to be  every other row or every other column of spins on the square lattice.

These sign conventions can be exactly expressed by a suitable choice of the variational parameters and can then be in principle learned during the variational optimization. However, we find in general more convenient to initialize our ansatz in one of the two sign conventions, and then optimize the resulting state. Our full variational ansatz then takes the form
\begin{equation} \label{eq: sign}
\ket{\Psi} = \sum_{\boldsymbol{\sigma}} (-1)^{M_{\mathcal{A}}(\boldsymbol{\sigma})} \Psi_{\mathrm{CNN}} (\boldsymbol{\sigma})  \ket{\boldsymbol{\sigma}} =  \sum_{\boldsymbol{\sigma}} \Psi^{\mathcal{A}}_{\mathrm{CNN}} (\boldsymbol{\sigma}) \ket{\boldsymbol{\sigma}}, 
\end{equation}
where $\Psi_{\mathrm{CNN}} (\boldsymbol{\sigma})$ is given by Eq.~\eqref{eq: bare CNN}. Since $\Psi_{\mathrm{CNN}} (\boldsymbol{\sigma})$ is complex-valued, the sign structure can in principle be changed by the network. However, the choice of the subset $\mathcal{A}$ does unavoidably present a bias in the variational ansatz and we unfortunately find that optimization is extremely challenging if an appropriate sign structure is not imposed. In this work, we perform an optimization with both of the limiting sign structures and pick the one which gives the better variational energy.

\subsection*{Enforcing $\mathcal{C}_{4}$ Symmetry}
In addition, as the model in Eq.~\eqref{eq: J1J2} is defined on a square lattice, we can expect that the ground state of the model transforms within an irreducible representation of the symmetry. However, while the CNN we use is explicitly translationally invariant, it is not explicitly $\mathcal{C}_{4}$ symmetric, i.e., t does not need to be an irreducible representation of the $\mathcal{C}_{4}$ group of fourfold rotations. As the $\mathcal{C}_{4}$ group is Abelian, the irreducible representations are one-dimensional. 

To ensure that the variational wavefunction is $\mathcal{C}_{4}$ symmetric, we symmetrize the wavefunction $\Psi_{\mathrm{CNN}}$. This can be done generically as follows: Denoting $\hat{c}_{4}$ to be the generator of the $\mathcal{C}_{4}$ group, the symmetrized wavefunction
\begin{equation}\label{eq: ansatz}
\tilde{\Psi}_{\mathrm{CNN}} (\boldsymbol{\sigma}) = \sum_{r=0}^{3} \omega^{r} \Psi^{\mathcal{A}}_{\mathrm{CNN}} (\hat{c}_{4}^{r} \boldsymbol{\sigma}),
\end{equation}
transforms within an irreducible representation with character $\omega$. Here, $\Psi^{\mathcal{A}}_{\mathrm{CNN}}$ is defined through Eq.~\eqref{eq: sign}. This symmetrization ensures that correlation functions have the correct spatially symmetry, a circumstance especially important in the striped order phase at large $J_{2}$, where the a-priori sign structure we start from is not rotationally invariant.

\begin{figure*}[t]
            \includegraphics[width=1.0\textwidth]{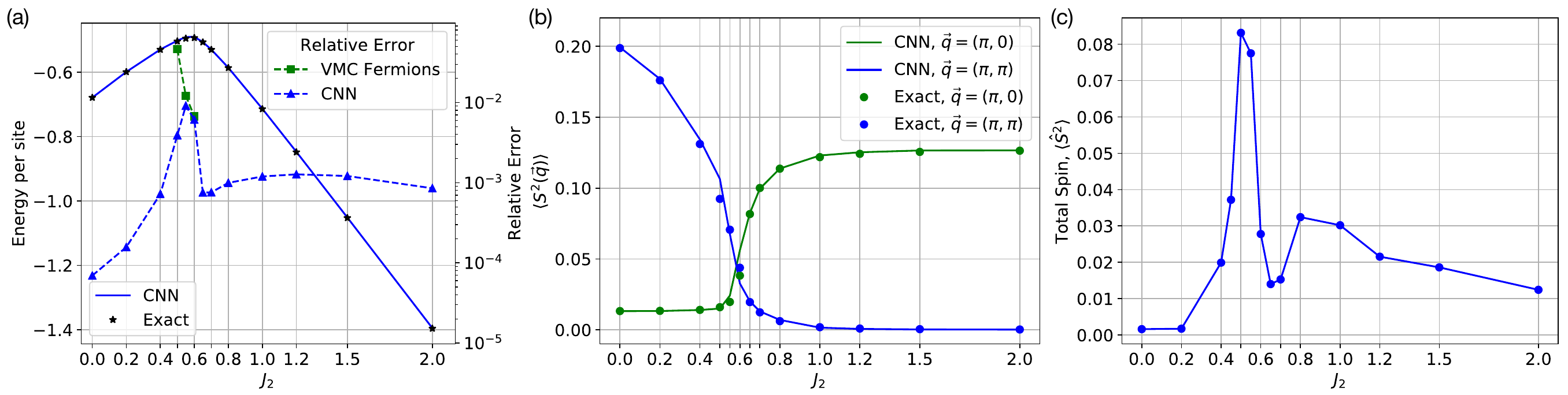}
    \caption{Simulation results on the $6 \times 6$ square lattice with periodic boundary conditions. (a) Energy comparison with ED results from Ref.~\cite{Schulz1996}. The CNN energies are indicated by the blue line while the black stars gives the exact values. The relative errors are plotted with respect to the right axis. The blue dashed line shows the relative error for our CNN ansatz and the green dashed line corresponds the the Gutzwiller-projected mean field fermionic variational ansatz from Ref.~\cite{WJHu2013}. (b) Spin-spin structure factor as defined in Eq.~\eqref{eq: spin correlation}. (c) The total spin $\langle \hat{\boldsymbol{S}}^{2} \rangle$, an extensive quantity. In the exact case, this value should be zero since the ground state is in the singlet sector.}
\label{fig: L=6}
\end{figure*}

\section{Variational Monte Carlo Optimisation}
In order to optimize the parameters of the variational ansatz Eq.~\eqref{eq: ansatz}, we use the method known as stochastic reconfiguration (SR) \cite{Sorella2007}, which can can be seen as an imaginary time evolution. 

Consider a variational wavefunction $\Psi(\lbrace \alpha^{0}_{k} \rbrace) \in \mathbb{C}^{2^n}$ which depends on a set of variational parameters $\lbrace \alpha^{0}_{k} \rbrace_{k=1, \dots, p}$. If we have a small variation in the parameters $\alpha_{k} = \alpha^{0}_{k} + \delta \alpha_{k}$, then the corresponding wavefunction can be written as
\begin{equation}
\Psi(\lbrace \alpha_{k} \rbrace) = \Psi(\lbrace \alpha_{k}^{0} \rbrace) + \sum^{p}_{k=1} \delta \alpha_{k} \mathcal{O}_{k} \Psi(\lbrace \alpha^{0}_{k} \rbrace),
\end{equation}
where 
$\mathcal{O}_{k} = \frac{\partial}{\partial \alpha_{k}} \log\left[\Psi(\lbrace \alpha^{0}_{k} \rbrace)\right]$
are the logarithmic derivatives.

The SR scheme is then essentially an imaginary time evolution, which is given to first order by
\begin{equation} \label{eqn: exp}
\Psi^{\prime}_{\mathrm{exact}} = (1 - \epsilon \hat{H}) \Psi.
\end{equation}
The optimal coefficients $\lbrace \delta \alpha^{0}_{k} \rbrace_{k=1, \dots, p}$ that minimize the distance to the new wavefunction $\Psi^{\prime}$ with respect to Fubini-Study metric,
\begin{equation}
\gamma(\phi, \Psi) = \arccos{\sqrt{\frac{\braket{\Psi|\phi} \braket{\phi|\Psi}}{\braket{\Psi|\Psi}\braket{\phi|\phi}}}},
\end{equation}
are then given by the solution to the linear equation
\begin{equation}\label{eq: SR}
\sum_{k^{\prime}} \left[ \langle \mathcal{O}_{k}^{\dagger} \mathcal{O}_{k^{\prime}} \rangle -  \langle\mathcal{O}_{k}^{\dagger}\rangle \langle\mathcal{O}_{k^{\prime}}\rangle\right]\delta\alpha_{k^{\prime}} = -\epsilon \left[ \langle\mathcal{O}_{k}^{\dagger} \hat{H}\rangle -  \langle\mathcal{O}_{k}^{\dagger}\rangle \langle\hat{H}\rangle \right].
\end{equation}
We update the parameters as $\alpha_{k} = \alpha^{0}_{k} + \delta \alpha_{k}$ and repeat the procedure until convergence is achieved.

Since each SR iteration requires solving the linear system, the computational complexity of each step is $\mathcal{O}(N_{w}^{2})$ (using the iterative conjugate gradients algorithm to solve the linear system), as compared to $\mathcal{O}(N_{w})$ for the stochastic gradient descent (SGD) method, where $N_{w}$ is the number of variational parameters. However, the SR method is known to perform better than SGD for variational optimization of small to mid-sized networks. 

The expectation values $\langle \cdots \rangle$ can be estimated using Monte Carlo sampling. For instance, the energy can be estimated as 
\begin{equation}
\begin{split}
\langle \hat{H} \rangle &= \frac{\sum_{\boldsymbol{\sigma}, \boldsymbol{\sigma}} \Psi^{*}(\boldsymbol{\sigma}') \bra{\boldsymbol{\sigma}}\hat{H}\ket{\boldsymbol{\sigma}'}\Psi(\boldsymbol{\sigma}) }{\sum_{\boldsymbol{\sigma}} \left| \Psi(\boldsymbol{\sigma})\right|^{2}} \\
&= \sum_{\boldsymbol{\sigma}} \left( \sum_{\boldsymbol{\sigma}'} \bra{\boldsymbol{\sigma}}\hat{H}\ket{\boldsymbol{\sigma}'} \frac{\Psi(\boldsymbol{\sigma}')}{\Psi(\boldsymbol{\sigma})} \right) \frac{ \left| \Psi(\boldsymbol{\sigma})\right|^{2}}{\sum_{\boldsymbol{\sigma}'} \left| \Psi(\boldsymbol{\sigma}')\right|^{2}} \\
&\approx \Big \langle \sum_{\boldsymbol{\sigma}'} \bra{\boldsymbol{\sigma}}\hat{H}\ket{\boldsymbol{\sigma}'} \frac{\Psi(\boldsymbol{\sigma}')}{\Psi(\boldsymbol{\sigma})}   \Big\rangle_{\mathrm{M}}
\end{split}
\end{equation}
where $\langle \cdots \rangle_{\mathrm{M}}$ denotes an average over a sample of configuration $\lbrace \boldsymbol{\sigma} \rbrace$ drawn from the probability distribution given by $|\Psi(\boldsymbol{\sigma})|^{2}$. The sample is easily obtained by the Metropolis algorithm \cite{Metropolis1953}. This average can be evalutated efficiently when the matrix $\bra{\boldsymbol{\sigma}}\hat{H}\ket{\boldsymbol{\sigma}'} $ is sparse which is the case for the Hamiltonian in Eq.~\eqref{eq: J1J2} when the basis corresponds to a tensor product of local spin degrees of freedom. Finally, the linear system Eq.~\eqref{eq: SR} is known to be highly ill conditioned especially for a network with numerous parameters. To alleviate this problem, we regularize the matrix by adding a multiple of the identity.

\begin{figure*}[t]
            \includegraphics[width=1.0\textwidth]{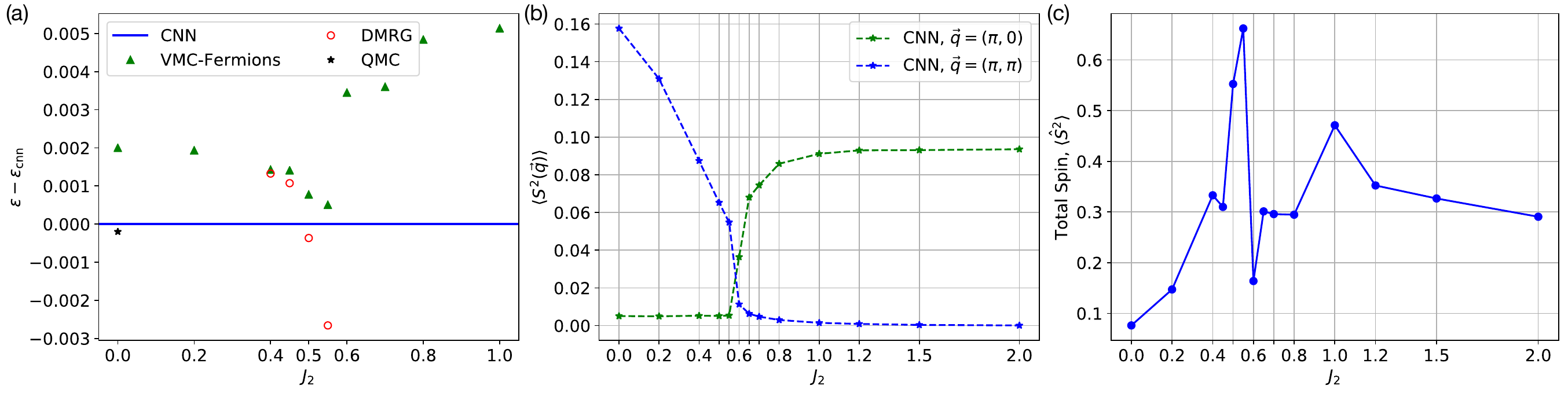}
    \caption{Simulation results on the $10 \times 10$ square lattice with periodic boundary conditions. (a) Energy comparison with other state of the art methods. We plot the energy difference per site with respect to the energy obtained from our CNN variational ansatz. The red open circles shows the values for the DMRG results of Ref.~\cite{SSGong2014} and the green triangles corresponds the the Gutzwiller-projected mean field fermionic variational ansatz. The black star shows the exact results from greens function quantum Monte Carlo for the sign-problem free point ($J_{2}=0$).  (b) Spin-spin structure factor as defined in Eq.~\ref{eq: spin correlation}. (c) The total spin $\langle \hat{\boldsymbol{S}}^{2} \rangle$. In the exact case, this value should be zero since the ground state is in the singlet sector.}
\label{fig: L=10}
\end{figure*}

\section{Results and Discussion}
Using the variational ansatz presented in Eq.~\eqref{eq: ansatz} together with the SR method described above, we now discuss the results obtained on the $6 \times 6$ as well as the $10 \times 10$ square lattice with periodic boundary conditions.
\subsection{Comparison with ED}
In the case of the $6 \times 6$ square lattice with periodic boundary conditions, the system is amenable to exact diagonalization (ED). In Fig.~\ref{fig: L=6}(a), we compare our variational energies with that of the lanczos ED results found in Ref.~\cite{Schulz1996}. We show also the relative error defined by $\left| \frac{E - E_{\mathrm{exact}}}{E_{\mathrm{exact}}}\right|$. While the relative error is of order $10^{-3}$ or lower for most of the parameter space, it is immediately clear that the simulation is the most challenging close to the point of maximum frustration $J_{2} \approx 0.5$. In our simulations, the largest relative error occurred at $J_{2} = 0.55$ which roughly corresponds to the point where our prior sign structure is maximally violated. This suggests that although the phases of the wavefunctions can be changed by the network, the prior sign structure presents a strong bias for the simulation.

To study the properties of the NQS beyond the energy expectation value alone, we also measure the spin-spin correlation structure factors defined by
\begin{equation} \label{eq: spin correlation}
S^{2}(\vec{q}) = \frac{1}{N(N+2)} \sum_{i,j} \langle \hat{\boldsymbol{S}}_{i} \cdot \hat{\boldsymbol{S}}_{j}\rangle e^{i \vec{q}\cdot(\vec{r}_{i} - \vec{r}_{j})},
\end{equation}
where $\vec{q}$ is the pitch vector and $N$ is the total number of spins in the system. For $J_{2} \lesssim 0.5$, the system is Neel ordered with pitch vector $\vec{q} = (\pi,\pi)$, while for $J_{2} \gtrsim 0.5$, the system has a stripe order with $\vec{q} = (\pi,0)$ or $(0,\pi)$. We plot these two structure factors in Fig.~\ref{fig: L=6}(b) in comparison with the exact values. The result is accurate for most of the parameter range apart from the transition region where frustration is maximal. 

Finally, we measure the total spin $\langle \hat{\boldsymbol{S}}^{2} \rangle$ of our variational wavefunction as shown in Fig.~\ref{fig: L=6}(c). The model has $\mathrm{SU}(2)$ symmetry and the ground state is known to be in the singlet sector where the total spin is zero. It is apparent from Fig.~\ref{fig: L=6} that the spike in the total spin coincides with the spike in relative error. As we have an even number of spin-1/2 degrees of freedom, the eigenvalues of the $\hat{\boldsymbol{S}}^{2}$ are of the form $s(s+1)$ with $s = 0,1,2,\dots ,N/2$ such that the next lowest eigenvalue is $2$. Since the expectation value of the total spin of our variational ansatz is much less than $2$ we can be certain that our wavefunction has a good overlap with the singlet sector.  

\subsection{Benchmarking with state of the art methods}
We now proceed to a system size which cannot be reached by ED calculations. The purpose is to benchmark our ansatz with the more established ones such as matrix product states (MPS) and traditional variational Monte Carlo wavefunctions. In Fig.~\ref{fig: L=10}(a), we compare our variational energies on the $10 \times 10$ cluster with the density matrix renormalization group results in Ref.~\cite{SSGong2014} [where $8192$ $\mathrm{SU}(2)$ states or equivalently $32000$ $\mathrm{U}(1)$ states were kept] as well as the Gutzwiller-projected mean field fermionic ansatz which is closely related to that used in Ref.~\cite{WJHu2013}. We see that the CNN ansatz has competitive energies in all the range of $J_2/J_1$. With the notable exception of the reported point at $J_2/J_1=0.55$, NQS energies are very close or better than those reported in the literature. 
This point is around the region where the prior sign structures are most violated, and indicates a residual, unoptimized sign structure as a likely source of systematic error. 
Nevertheless, it is encouraging that by using a relatively simple ansatz, without much prior input information about the physics of the system, one is able to achieve variational energies competitive with, and in most of the parameter space better than, other state of the art methods. In the Appendix, we provide a table containing all the variational energies obtained.
A further possible origin for the non-optimal performance at $J_2/J_1=0.55$ can also be traced in symmetry violations in our CNN ansatz. Most notably, whereas the fermionic ansatz as well as the DMRG calculations preserve $\textrm{SU}(2)$ symmetry explicitly, the CNN does not. This can be seen in Fig.~\ref{fig: L=10}(c), where the total spin of our variational form is substantially different from zero, albeit still being less than $2$ (the next lowest eigenvalue of the total spin operator $\hat{\boldsymbol{S}}^{2}$) and hence ensuring a decent overlap with the singlet sector. Despite the mixing with other spin sectors, the features of the Neel and striped magnetic orders can still be seen in the spin-spin correlations in Fig.~\ref{fig: L=10}(b).

Before concluding the section, we would like to mention also a previous work Ref.~\cite{Liang2018}, which used a slightly different CNN as a variational ansatz to study the same problem. On the $10\times 10$ cluster with $J_{2}=0.5$, a variational energy per site of $-0.4736$ was obtained as compared to our result $-0.4952$. This differs significantly from other state of the art methods as can be seen from the scale on the energy axis of Fig.~\ref{fig: L=10}(a). There are some qualitative differences between their ansatz and ours: 1) Ref.~\cite{Liang2018} used real parameters while ours are complex-valued. 2) The non-linearity in Ref.~\cite{Liang2018} is introduced via max-pooling operations while we use non-linear activation functions. In addition, we symmetrize our ansatz to have $\mathcal{C}_{4}$ rotation symmetry and also provided an initial prior sign structure. 3) The optimization technique used in Ref.~\cite{Liang2018} is different from the SR method we have employed. 

\subsection{Discussion}
The results obtained in this paper provide a tangible evidence that NQS are a competitive variational ansatz to study challenging open problems such as frustrated magnets.  
While we have provided here direct numerical evidence for their suitability to study the stability of the spin liquid phase in the $J_1$-$J_2$ model, some open aspects 
have not been yet addressed in this work, and will be the focus of future research. 

First, since the system is gapless, finite size effect are large such that accurate extrapolations to the thermodynamic limit are necessary. 
A more computationally demanding simulation campaign would be required to provide a firm finite-size extrapolation of the magnetic correlations presented here. 

Second, the networks we have used here are comparably much smaller, in terms of depth and number of trainable parameters, than state-of-the-art models used in modern deep learning applications. Our networks contain at least three orders of magnitude less parameters than what found in typical deep CNNs routinely used for image recognition, and the margin for future improvements therefore seems quite substantial. In order to bridge this gap, more expressive models could be adopted, for example along the lines of the recently introduced auto-regressive models~\cite{SharirDeep2019} for quantum states. 
Finally, it is rather clear from our simulations that the performance of our ansatz is also linked to the sign structure of the many-body state. This implies that ``learning" the correct sign structure within the parametrization we adopted presents a challenge for the SR optimization we use. In this context, it would be interesting to see how modern machine learning techniques such as reinforcement learning can help tackle this problem.

\section*{Acknowledgments}
We gratefully acknowledge Francesco Ferrari and Federico Becca for helpful discussions and providing VMC energies for the Gutzwiller-projected mean field fermionic wavefunctions. 
We also acknowledge discussions with Juan Carrasquilla and Yusuke Nomura. KC was supported by the European Unions Horizon 2020 research and innovation program (ERC-StG-Neupert-757867-PARATOP). KC thanks the Flatiron Institute founded by the Simons foundation for computational resources. The ANN computations were based on NetKet \cite{NetKet}.

\section*{Appendix: Variational Energies}
In the table below we show the exact values (including error bars) for the variational energies obtained in this work.
\begin{widetext}

\begin{center}
\centering
\begin{table}[h]
\begin{tabular}{|c|c|c|c|c|c|c|c|c|c|}
\hline
$6\times 6$  & $J_{2} = 0.0$ &  $J_{2} = 0.2$ &$J_{2} = 0.4$ & $J_{2}= 0.45$ &  $J_{2}=0.5$  & $J_{2}=0.55$  & $J_{2}=0.6$ & $J_{2}=0.8$ & $J_{2}=1.0$  \\ \hline
Exact & $-0.678872$ & $-0.599046$ &$-0.529745$   & - & $-0.503810$ & $-0.495178$ & $-0.493239$ & $-0.586487$ & $-0.714360$\\ \hline 
VMC & - & - &$-0.52715(1) $ & $-0.51364(1)$ & $-0.50117(1)$ & $-0.48992(1)$ & - & - & - \\ \hline  
DMRG & - & - &$-0.529744$   & $-0.515655$ & $-0.503805$ & $-0.495167 $ & - & - & -\\ \hline  
CNN & $-0.67882(1)$ & $-0.59895(1)$ & $-0.52936(1)$   & $-0.51452(1)$ & $-0.50185(1)$ & $-0.49067(2)$ & $ -0.49023(1)$ & $ -0.58590(1)$ & $-0.71351(1)$ \\ \hline  
\hline
$10\times 10$  & $J_{2} = 0.0$ &  $J_{2} = 0.2$ &$J_{2} = 0.4$ & $J_{2}= 0.45$ &  $J_{2}=0.5$  & $J_{2}=0.55$  & $J_{2}=0.6$ & $J_{2}=0.8$ & $J_{2}=1.0$  \\ \hline
VMC & $-0.66935(1)$& $-0.59082(1)$ &$-0.52188(1)$ & $-0.50811(1)$ & $-0.49521(1)$ & $-0.48335(1)$ & $-0.47259(1)$ & $-0.56899(1)$ & $-0.69123(1)$ \\ \hline  
DMRG & - & - &$-0.522391$   & $-0.507976$ & $-0.495530$ & $-0.485434$ & - & - & -\\ \hline  
CNN & $-0.67135(1)$ & $-0.59275(1)$ & $-0.52371(1)$   & $-0.50905(1)$ & $-0.49516(1)$ & $-0.48277(1)$ & $-0.47604(1)$ & $-0.57383(1)$ & $-0.69636(1)$ \\ \hline  
\end{tabular}
\caption{Comparison between Gutzwiller-projected mean field fermionic wavefunctions VMC \cite{WJHu2013} (on the $10\times10$ case the energies were provided by F. Ferrari and F. Becca), DMRG with $8192$ $\mathrm{SU}(2)$ states or equivalently $32000$ $\mathrm{U}(1)$ states \cite{SSGong2014} and the CNN used in this paper. The exact energies on the $6\times 6$ case were take from Ref.~\cite{Schulz1996}.}
\end{table}
\end{center}

\end{widetext}

\bibliography{biblio}

\clearpage
\end{document}